\documentclass[twocolumn,amsmath,amssymb,aps,]{revtex4-1}
\usepackage{bm}
\usepackage[colorlinks=true,urlcolor=blue,linkcolor=blue,citecolor=blue]{hyperref}
\usepackage{color}
\usepackage{graphics}
\usepackage{graphicx}
\usepackage{epsfig}
\usepackage{amssymb}
\usepackage{amsmath}
\usepackage{hyperref}
\usepackage{physics}

\usepackage{array}                 

\usepackage{amsfonts}           

\begin{document}


\title{Formation of Two-Ion Crystals by Injection from a Paul-Trap Source into a High-Magnetic-Field Penning Trap}

\author{J.~Berrocal$^{1}$} 
\author{E.~Altozano$^1$}
\author{F.~Dom\'inguez$^1$}
\author{M.J.~Guti\'errez$^{1}$}\thanks{Present address: GSI Helmholtzzentrum für Schwerionenforschung GmbH, D-64291, Darmstadt, Germany.}
\author{J.~Cerrillo$^2$}
\author{F.J.~Fern\'andez$^3$}
\author{M.~Block$^{4,5,6}$}
\author{C.~Ospelkaus$^{7,8}$}
\author{D.~Rodr\'iguez$^{1,9}$}\email[]{danielrodriguez@ugr.es}
\affiliation{
$^1$Departamento de F\'isica At\'omica, Molecular y Nuclear, Universidad de Granada, 18071 Granada, Spain \\
$^2$\'Area de F\'isica Aplicada, Universidad Polit\'ecnica de Cartagena, 30202 Cartagena, Spain\\
$^3$Departamento de Arquitectura y Tecnolog\'ia de Computadores, Universidad de Granada, 18071 Granada, Spain\\
$^4$Department Chemie - Standort TRIGA, Johannes Gutenberg-Universit\"at Mainz, D-55099, Mainz, Germany \\
$^5$GSI Helmholtzzentrum f\"ur Schwerionenforschung GmbH, D-64291, Darmstadt, Germany \\
$^6$Helmholtz-Institut Mainz, D-55099, Mainz, Germany \\
$^7$Institute of Quantum Optics, Leibniz Universit\"at Hannover, Welfengarten 1, 30167 Hannover, Germany\\
$^8$Physikalisch-Technische Bundesanstalt, Bundesallee 100, 38116 Braunschweig, Germany\\
$^9$Centro de Investigaci\'on en Tecnolog\'ias de la Informaci\'on y las Comunicaciones, Universidad de Granada, 18071 Granada, Spain
}

\date{\today}

\begin{abstract}
Two-ion crystals constitute a platform for investigations of quantum nature that can be extended to any ion species or charged particle provided one of the ions in the crystal can be directly laser-cooled and manipulated with laser radiation. This paper presents the formation of two-ion crystals for quantum metrology in a 7-tesla open-ring Penning trap. $^{40}$Ca$^+$ ions are produced either internally by photoionization or externally in a (Paul-trap) source, transported through the strong magnetic field gradient of the superconducting solenoid, and captured in-flight with a mean kinetic energy of a few electronvolts with respect to the minimum of the Penning-trap potential well. Laser cooling of the two-ion crystal in a strong magnetic field towards reaching the quantum regime is also presented with particular emphasis on the cooling of the radial modes.
\end{abstract}


\maketitle

\section{Introduction}
The precise control of individual (or a pair of) charged particles, or particle-antiparticle pairs in Penning traps is a pre-requisite to perform accurate measurements of the eigenfrequencies of the stored particle and/or antiparticle \cite{Corn1992,Rain2004,Stur2014,Ulme2015} with implications in tests of fundamental symmetries (see e.g. Ref.~\cite{Myer2019} for a recent review). The control of the charged particle down to the lowest motional energy in a Penning trap, only possible by performing laser cooling \cite{Good2016,Hrmo2019}, moves the experiments on precise motional frequency measurements to a new regime of quantum nature \cite{Cerr2021}. However, in most cases the cooling of the charged particle under study relies on the Coulomb interaction between the particle and an auxiliary ion that can be laser cooled \cite{Schm2005}. The ion (sensor) and the charged particle to be studied may be stored in physically separated Penning traps \cite{Hein1990,Rodr2012,Corn2016,Bohm2021}, in different potential wells \cite{Corn2021}, or in the same one \cite{Guti2019b,Cerr2021}. The sensor shall provide the information of the system as the target is blind to laser radiation. If such a platform can be coupled to an external ion source, it can turn into a universal atomic or molecular mass spectrometer. This is crucial for the production of any target ion as well as the sensor species in well-controlled manner, outside the Penning trap. For example, the formation of antimatter particles or exotic radionuclides is usually accomplished in nuclear reactions resulting in few particles at high energy and with large momentum spread, so that an additional preparation stage is mandatory \cite{Gabr1999,Smor2017,Blau2013}. Furthermore, the production of the sensor ion outside the Penning trap will avoid charge exchange between the laser-cooled sensor ion/s and hotter (sensor) atoms \cite{Luca2004} to the advantage of experiments where the coolant source has to be in continuous operation. The injection of ions produced outside a Penning trap is well-known in experiments devoted to precision mass spectrometry, where the target ion is probed without laser cooling (see e.g. extreme scenarios like described in Refs.~\cite{Bloc2010,Mina2012}). This is not the case in laser cooling Penning-trap experiments (and in general in similar experiments with other kinds of ion traps) devoted to quantum sensing or quantum information processing \cite{Gilm2017,Jord2019,Hrmo2019}.  In those experiments, the ion species to be laser-cooled are generally produced inside the Penning trap either by laser-desorption on a metallic sample located close to the trap electrodes (see e.g. \cite{Niem2019}), or by photoionization or electron impact ionization of atoms released from an oven as initially done at the experiments reported here. Thus, the combination of external ion production and in-trap laser cooling has not been much exploited. Results from only a few experiments have been reported. In Ref.~\cite{Bake2021}, Be$^+$ ions were produced by laser ablation upstream from a Penning trap, where they were injected, laser cooled and utilized for sympathetic cooling of positrons.
\begin{figure*}[t!]
\centering\includegraphics[width=0.9\linewidth]{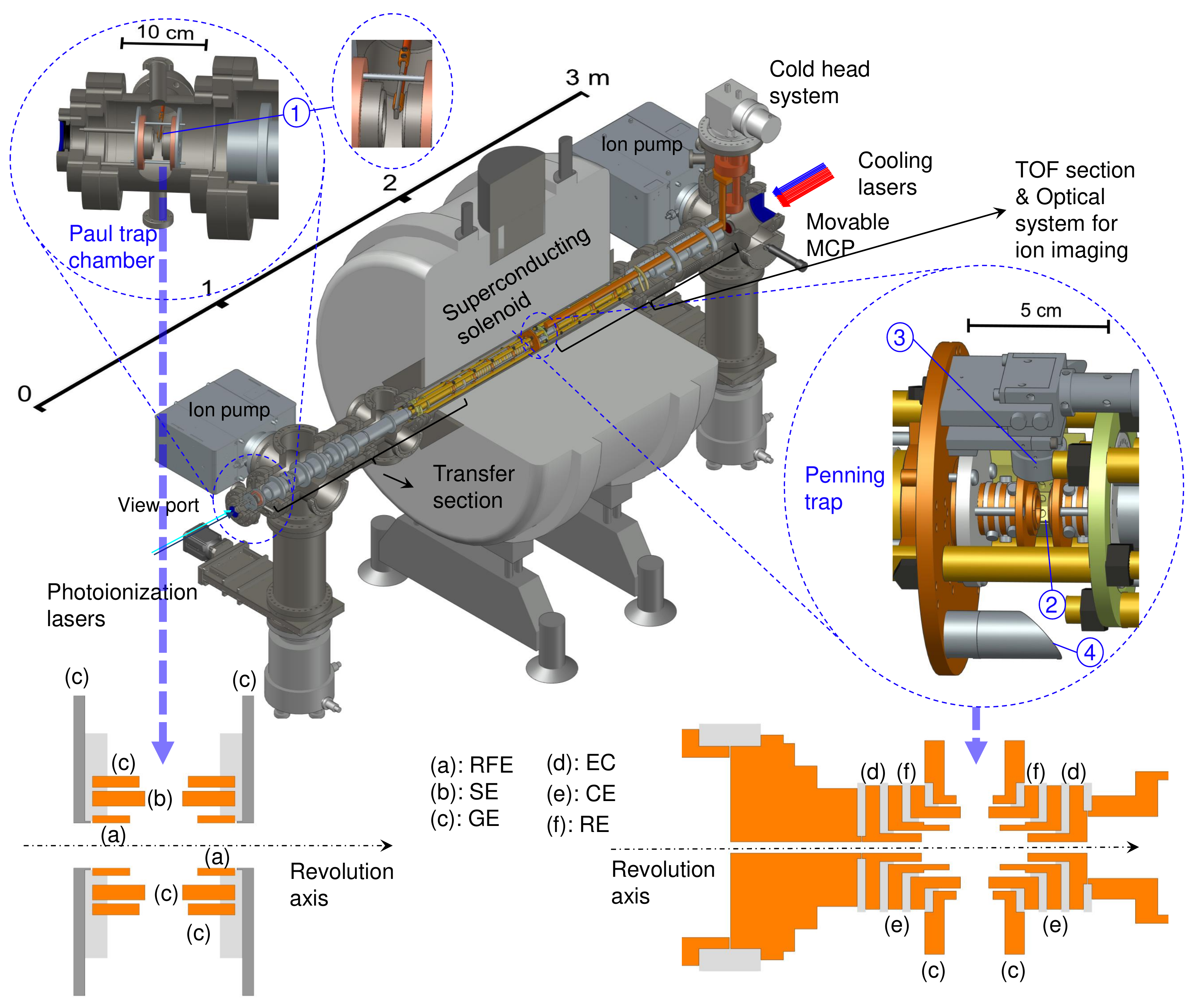}
\vspace{-0.4cm}
\caption{3D CAD drawing of the Penning-traps beamline. The most important elements are indicated. The open-ring traps (Paul  and Penning) are zoomed. Number 1 indicates an atomic oven in the Paul trap chamber, number 2 points out the MACOR structure surrounding the atomic oven oriented to the center of the open-ring Penning trap. Number 3 indicates the housing of the optical objective to collimate the fluorescence photons and number 4 points at the support structure of one of the mirrors to direct the cooling laser beams in the radial plane of the ions' motion. Longitudinal cuts of the Paul and Penning trap are shown indicating the different electrodes in orange (insulators are in grey). RFE, stands for radiofrequency electrode(s), SE for switching electrode and GE for all the electrodes at ground. The nomenclature for the Penning-trap electrodes is as in Ref~\cite{Guti2019}: EC, CE and RE are the endcap(s), correction(s), and ring(s) electrodes, respectively.  More details are given in the text.} \label{trapsensor_beamline}
\end{figure*}
Ion crystals in a Penning trap from externally produced $^{24}$Mg$^+$ ions have been formed at GSI-Darmstadt in the SpecTrap experiment \cite{Murb2016} aiming at creating a bath of coolant ions for sympathetic cooling of highly-charged ions, to perform laser spectroscopy experiments on the latter. In that work, only axial cooling was applied and crystals formed with many ions were reported showing different planes in the radial direction, but never a few-ion crystal oriented in the direction of the magnetic field. Other experiments using externally produced target ions are those on quantum-logic spectroscopy on highly-charged ions at PTB-Braunschweig \cite{Leop2019} or to perform studies on $^{232}$Th$^+$ at the University of Mainz \cite{Groo2019}. Both use radiofrequency traps and still work with the internal ion production of the coolant-and-sensor ion.

In this publication we report on the formation of ion crystals in an open-ring Penning trap \cite{Guti2019}, when the ions are produced by photoionization in a Paul trap made of concentric rings \cite{Corn2015,Domi2017}. This trap is located in the beamline outside the superconducting solenoid providing the largest magnetic field used until now for a Penning-trap laser-cooling experiment \cite{Guti2019}. The injection can be controlled down to just one ion by means of evaporative cooling by lowering down the trap voltages. In addition, the geometry of the Paul trap with an open diameter of 10~mm allows through-transport of any ion species produced by other sources upstream, which can be also relevant for accelerator-based experiments. The production of a single laser-cooled ion and a balanced two-ion crystal in a 7-tesla magnetic field are demonstrated and characterized. The two mechanisms utilized to cool the magnetron motion and motional modes in both systems have been studied and the advantage of having a high magnetic field strength is presented. In Appendix~\ref{coulomb} we show Coulomb crystals with large number of ions produced with this setup, that we use to obtain an upper limit of the ions' temperature. Coulomb crystals have previously been extensively studied by several groups (see e.g. \cite{Boll1993,Mava2013}).

\section{Experimental setup} \label{sectwo}

The Penning-traps beamline at the University of Granada is shown in Fig.~\ref{trapsensor_beamline}. It has been modified from that described in Ref.~\cite{Guti2019} by introducing a Paul trap \cite{Corn2015} in the transfer section. The superconducting solenoid provides in the room temperature bore two highly-homogeneous magnetic field regions in a volume of 1~cm$^3$ each, separated by 20~cm. The center of the first of these regions ($\Delta B/B\sim 10$~ppm) coincides with the center of a preparation Penning trap made of a stack of cylinders \cite{Corn2016b}, which is grounded in the experiments described here and it is foreseen for further preparation of other ion species. The second highly-homogeneous region ($\Delta B/B\sim 0.1$~ppm) coincides with the center of the open-ring Penning trap displayed in Fig.~\ref{trapsensor_beamline} \cite{Guti2019}, where the experiments on the two-ion crystal are carried out. The trap-tower is housed in a customized tube inside the magnet bore. Two turbomolecular pumps, one at each side of the magnet, with a pumping speed of 800~l/s (for N$_2$) each, and two ion pumps, clearly visible in the figure with pumping speeds of 300~l/s and 600~l/s at the transfer and time-of-flight (TOF) sections, respectively, provide a vacuum at the pump side in the order of $10^{-10}$~mbar, which allows reaching about $10^{-9}$~mbar in the trap volume. This is possible by means of a copper bar which ends close to the trapping region and it is connected on the other side to the first stage of a cold-head system at 40~K. The two traps are zoomed in Fig.~\ref{trapsensor_beamline}, where also the longitudinal cuts with the different electrodes are shown.

\begin{figure}[t]
\centering\includegraphics[width=1.03\linewidth]{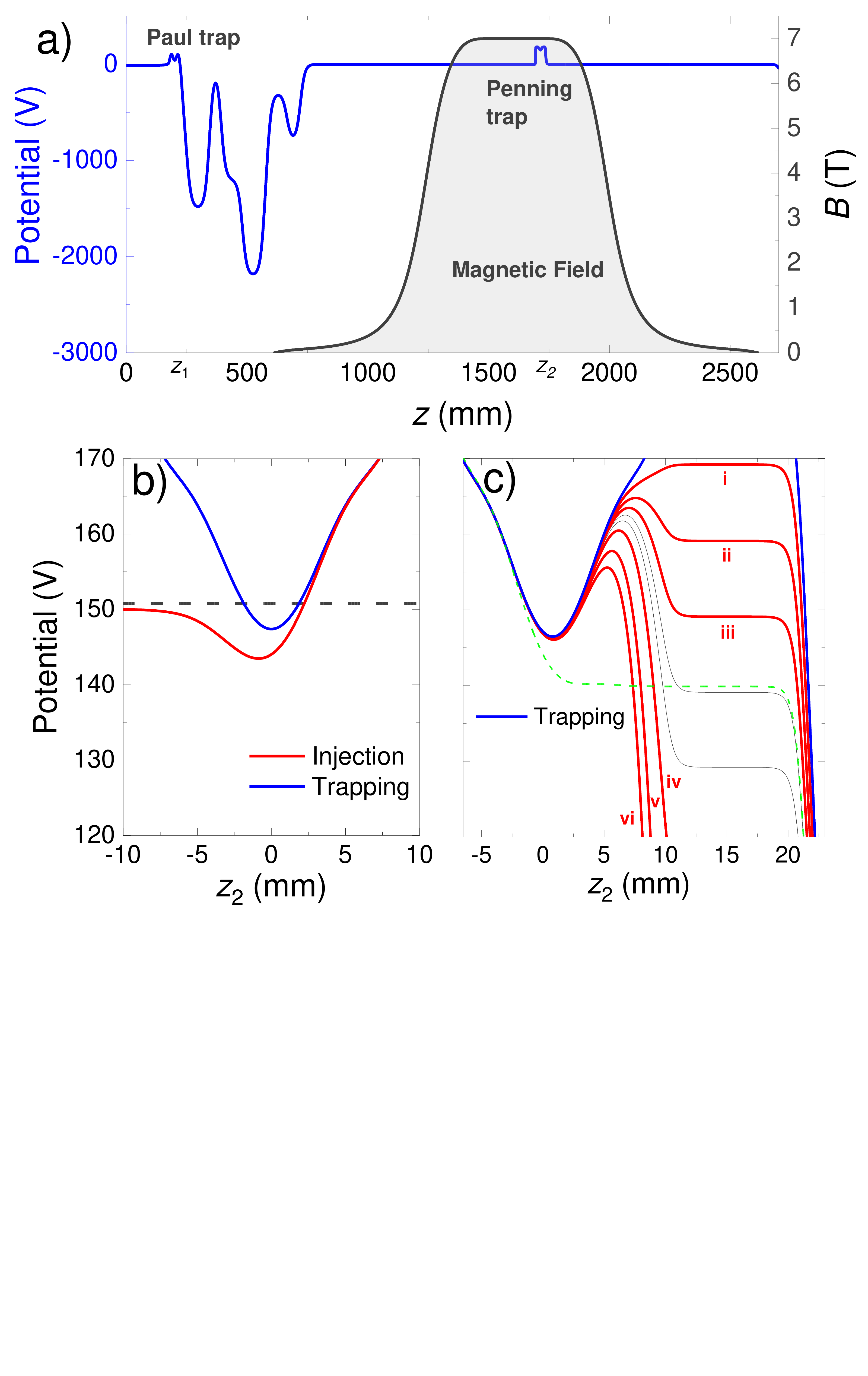}
\vspace{-5.4cm}
\caption{a) Electric potentials and magnetic field strength along the Penning-traps beamline. The electrostatic potential is given on the left axis while the magnetic field strength on the right one. $z_1=z-200$~mm and $z_2=z-1733$~mm. b) DC potential in the open-ring Penning trap for injection and trapping, and c) potential shapes for adiabatic and energy-selective extraction of the ions from the Penning trap (not all the potentials used in the experiment are shown); (i) $V_{\scriptsize{\hbox{EC}}}=170$~V,  (ii) $V_{\scriptsize{\hbox{EC}}}=160$~V, (iii) $V_{\scriptsize{\hbox{EC}}}=150$~V, (iv) $V_{\scriptsize{\hbox{EC}}}=110$~V, (v) $V_{\scriptsize{\hbox{EC}}}=50$~V, and (vi) $V_{\scriptsize{\hbox{EC}}}=-20$~V. $V_{\scriptsize{\hbox{CE}}}=168$~V and $V_{\scriptsize{\hbox{RE}}}=150$~V. The dashed-grey line in panel b) represents the mean ions' kinetic energy. The dashed-green line in panel c) is a potential configuration to extract all the ions from the trap. The DC potentials are extracted from SIMION simulations.}\label{fig:pot_line}
\end{figure}

\begin{figure}[t]
\centering\includegraphics[width=1.0\linewidth]{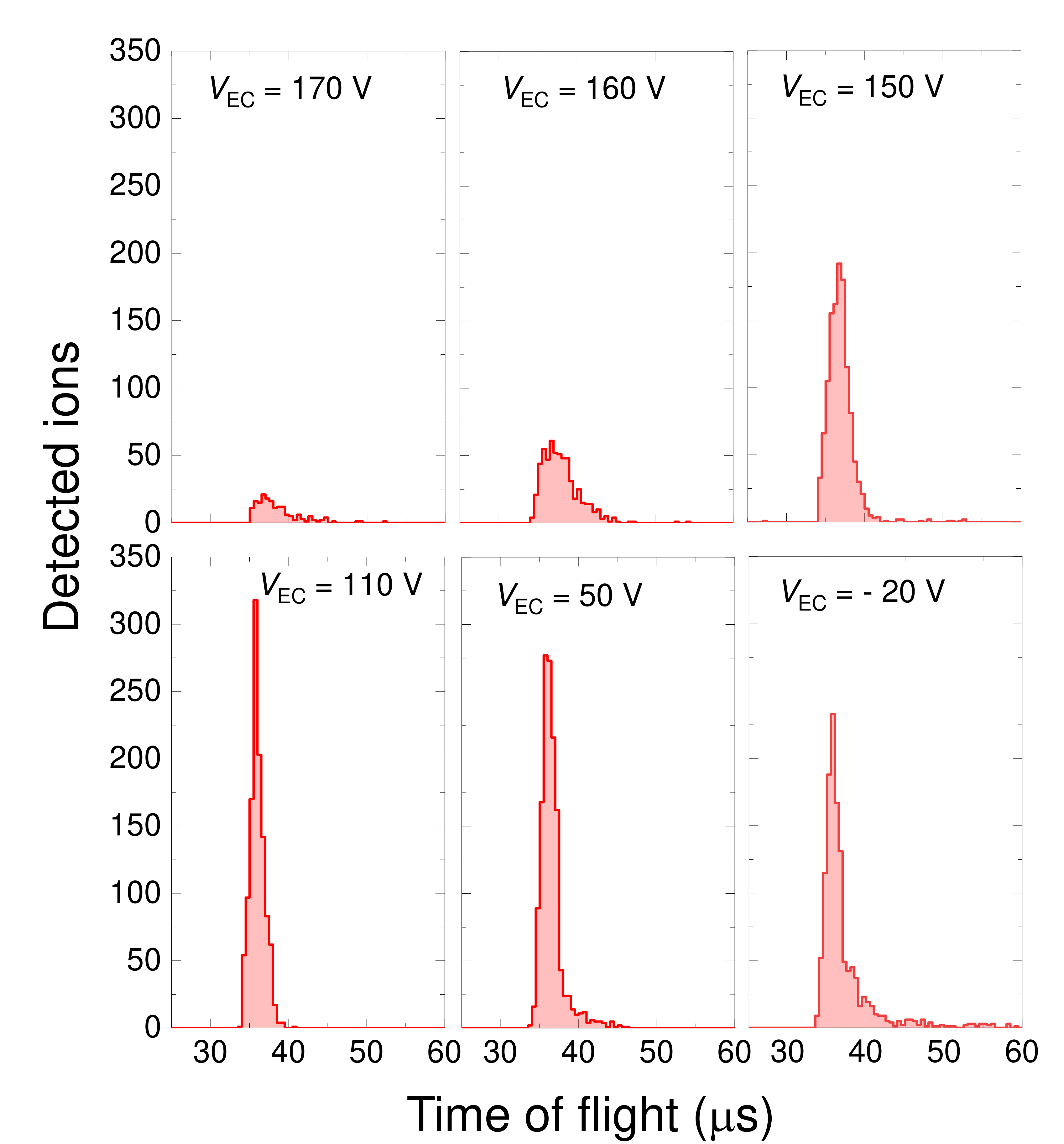}
\vspace{-0.5cm}
\caption{Time-of-flight signal of ions from the open-ring Penning trap to the MCP detector for several values of $V_{\scriptsize{\hbox{EC}}}$. 1000~cycles have been accumulated. The ions are produced by photoionization in  the Paul trap during one second, ejected, transferred, and captured in the Penning trap, held  for 20~ms, and ejected again towards the MCP detector at the end of the beamline. Different potentials have been applied for extraction. These potentials are depicted in Fig.~\ref{fig:pot_line}c from (i) to (vi). The ions are not cooled for these measurements.}\label{fig:tof}
\end{figure}

The production of ions is done by photoionization of calcium atoms, using two laser beams with wavelengths around $423$~nm (tunable) and 375~nm (free running). The laser beams for both internal and external production are the same (photoionization lasers in Fig.~\ref{trapsensor_beamline}). The laser for the degenerate $^1$S$_0\rightarrow ^1$P$_1$ transition (423~nm) has to be tuned differently for the production inside the Paul trap ($B=0$~tesla) and inside the Penning trap ($B=7$~tesla). For the latter, the non-degenerate transition $^1$S$_0\rightarrow ^1$P$_{1,+1}$ is chosen. 

\begin{figure}[t]
\centering\includegraphics[width=0.95\linewidth]{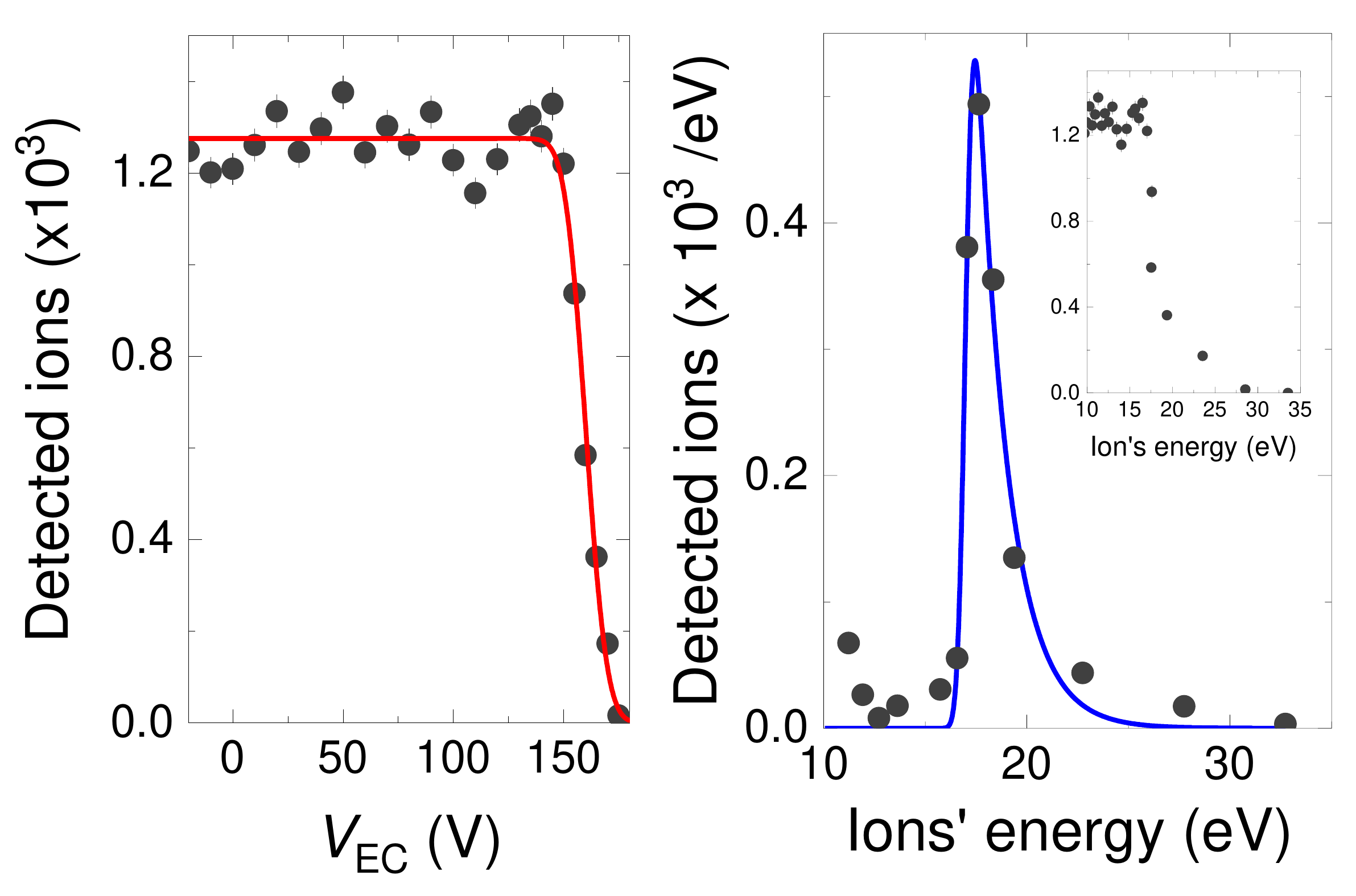}
\vspace{-0.4cm}
\caption{Energy spread of the ions in the Penning trap before cooling. The data points in the left panel are the number of detected ions with time-of-flight between 30 and 60~$\mu$s for different values of $V_{\scriptsize{\hbox{EC}}}$ (some of them shown in Fig.~\ref{fig:pot_line}c). The right panel shows the data points from the derivative of the number of detected ions as a function of the trapped-ion's energy, shown in the inset. The blue-solid line is the fit using an exponentially modified Gaussian distribution.}\label{fig:spread}
\end{figure}

Figure~\ref{fig:pot_line}a shows the electrostatic potential and magnetic field along the beamline in the axial direction. The Paul trap \cite{Corn2015} is operated for these experiments at a radiofrequency $\omega _{\scriptsize{\hbox{RF}}}=2\pi \times 600$~kHz with $V_{\scriptsize{\hbox{RF}}}=230$~V$_{\hbox{\scriptsize{pp}}}$. This corresponds to $q_z$ and $q_r$ values of $\approx 0.49$ and $0.25$, respectively.  
In the extraction process, the RF field is switched off with a decay time constant of the order of a microsecond, and one of the outer rings in the Paul trap (SE in Fig.~\ref{trapsensor_beamline}) is pulsed providing the DC potential shape in the axial direction to push the ions towards the Penning trap. The RF phase is not locked during the extraction. The measured ions' kinetic energy is centered at 150.8(7)~eV (dashed line in Fig.~\ref{fig:pot_line}b), with a standard deviation of 10.4~eV. A movable micro-channel plate (MCP) detector is located at the end of the beamline for diagnosis. The mean time-of-flight of the $^{40}$Ca$^+$ ions from the Paul trap to the MCP detector at the end of the beamline is 83.9~$\mu$s with $\Delta t_{\hbox{\scriptsize{FWHM}}}=5.5$~$\mu$s. The time-of-flight from the Paul trap to the Penning trap is centered at $47$~$\mu$s. The voltage configurations for injection and trapping in the open-ring Penning trap are depicted in Fig.~\ref{fig:pot_line}b. The trapping efficiency is $\approx 30$\%. It is obtained from the ratio between the detected ions ejected from the Penning trap and the detected ions in transmission. The voltages applied for trapping are $V_{\scriptsize{\hbox{EC}}}=180$~V, $V_{\scriptsize{\hbox{CE}}}=168$~V and $V_{\scriptsize{\hbox{RE}}}=150$~V. The ions are extracted from the open-ring Penning trap under different configurations, i.e., by applying different voltages to the endcap electrode $V_{\scriptsize{\hbox{EC}}}$. Some of these voltages are shown in Fig.~\ref{fig:pot_line}c. Figure~\ref{fig:tof} shows time-of-flight distributions for the configurations marked from (i) to (vi) in Fig.~\ref{fig:pot_line}c. These measurements serve to obtain the energy spread of the trapped ions. The left panel of Fig.~\ref{fig:spread} shows the number of detected ions versus $V_{\scriptsize{\hbox{EC}}}$. This number does not increase when the potential barrier decreases from a certain level. Only when the trap is fully open (dashed-green line in Fig.~\ref{fig:pot_line}c), the count rate increases by 14(2)\%, due to low energy ions. The number of detected ions is shown as a function of trapped-ions' energy in the inset of the right panel after considering the simulation results in Fig.~\ref{fig:pot_line}c. The right panel is obtained from the derivative of the data in the inset. An exponentially modified Gaussian distribution is used for the fit. $E=17.0(1)$~eV with a standard deviation of 1.5~eV is obtained for the trapped ions.

\begin{figure}[b]
\centering\includegraphics[width=1.0\linewidth]{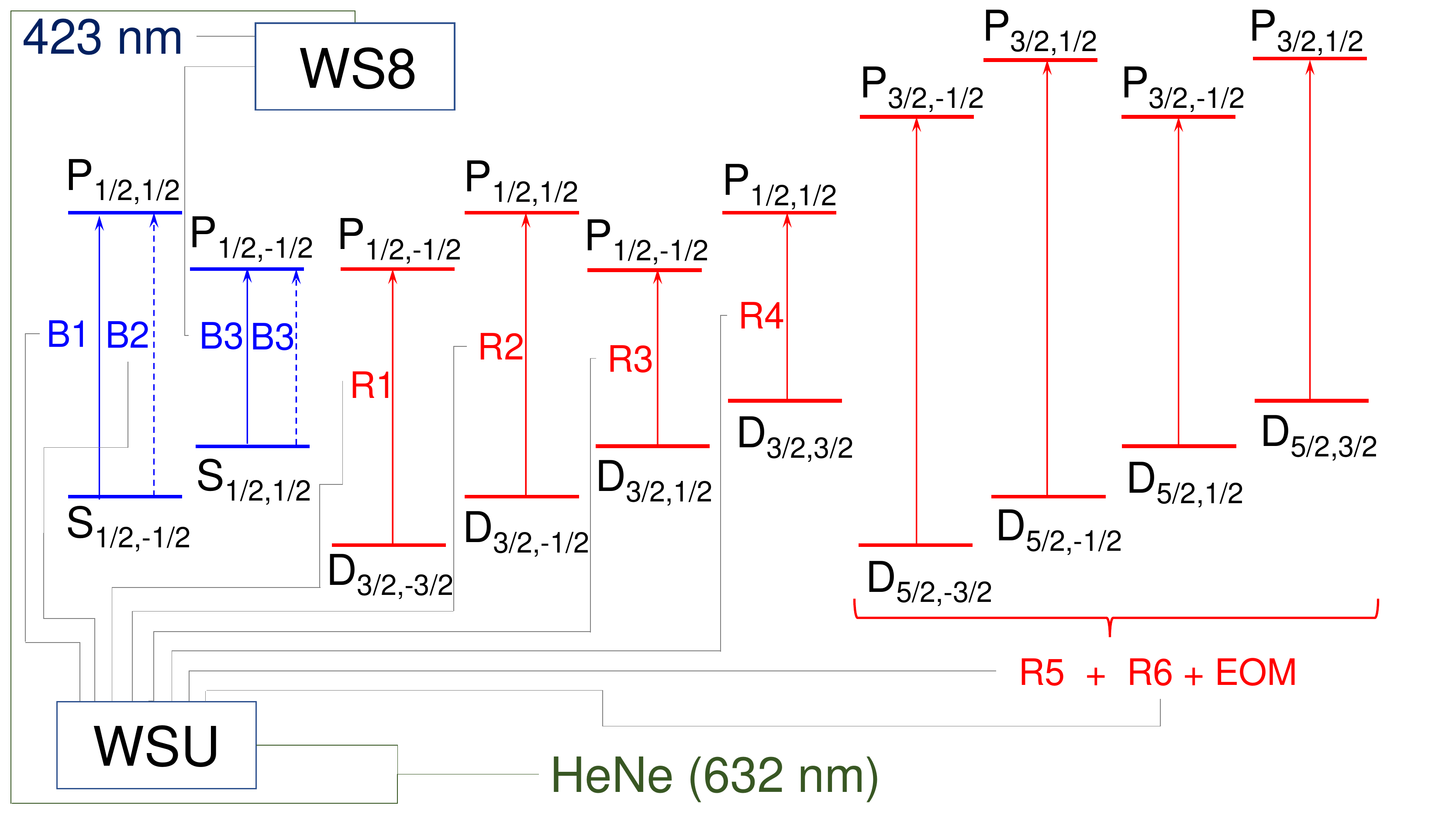}
\vspace{-0.5cm}
\caption{Driven transitions in the $^{40}$Ca$^+$ ion to perform Doppler cooling in 7~tesla \cite{Guti2019}. Two wavemeters are used to regulate the nine laser frequencies needed in the experiment for cooling (B1-B3) and pumping the dark states (R1-R6), and the 423-nm laser for photoionization. See text for further details.}\label{fig:lasers}
\end{figure}

The transitions for Doppler cooling in 7~tesla are shown in Fig.~\ref{fig:lasers}, following the  scheme presented in Ref.~\cite{Guti2019}. In that publication, only cooling in the axial direction was applied (solid arrows in Fig.~\ref{fig:lasers}). Radial cooling \cite{Itan1982} (dashed arrows) was introduced later and applied with and without axialization for the experiments presented here (for axialization see e.g. \cite{Hrmo2019}). The laser system comprises one free-running diode laser ($\lambda =375$~nm) and ten tunable diode lasers emitting at $\lambda =397$~nm (B1, B2 and B3), $\lambda =866$~nm (R1, R2, R3 and R4), $\lambda =854$~nm (R5 and R6) and $\lambda =423$~nm, stabilized by means of two wavelength meters (WSU and WS8 from High-finesse), as shown in Fig.~\ref{fig:lasers}, both with absolute accuracies of $10$~MHz, and calibrated by means of a HeNe laser every 180~seconds. The standard deviation of the laser frequencies while running the experiment is in the range 400-700~kHz for B1, B2 and B3. The laser powers have been varied in the different measurements. For the single ion and two-ion crystal these were about 520~$\mu$W (B1), 340~$\mu$W (B2), 280~$\mu$W (B3$_{\hbox{\scriptsize{axial}}}$) and 1.6~mW (B3$_{\hbox{\scriptsize{radial}}}$). The laser powers for R1-R6 range from $380$~$\mu$W to $1.75$~mW. We define the beam diameter as the distance from $I_{max}$ down to $I_{max}/e^2$ in the beam profile monitor. For the cooling-laser beams in the radial direction the diameters are measured externally at distances from the laser outputs equal to the distances to the trap center resulting in 320~$\mu$m (B2) and 220~$\mu$m (B3). In the axial direction the values amount to 500~$\mu$m for B1 and B3 and  about 1.6~mm for the IR lasers. 
The experiment is running with all laser beams above saturation.  An estimation of the ion's temperature has been obtained from plasma considerations in large crystals (Appendix~\ref{coulomb}).

\section{Penning trap results: single ion and two-ion crystals}
The motion of a single ion with mass $m$ and electronic charge $q$ in a Penning trap with a magnetic field $\vec B=B\vec k$, is the superposition of three eigenmotions \cite{Brow1986}, one in the axial direction defined by the unitary vector $\vec k$, with characteristic frequency $\omega _z$ independent of the magnetic field, and two in the radial plane with frequencies $\omega _{c'}$ and $\omega _m$, fulfilling the relationship
\begin{equation}
\omega _{c'}^2+\omega _z^2 +\omega _m ^2=\omega _c^2,
\end{equation}
where 
\begin{equation}
\omega _c =\frac{q}{m}B
\end{equation}
is the cyclotron frequency.  $\omega _{c'}=\omega _c/2 +\omega _1$ and $\omega _m=\omega _c/2 -\omega _1$, with
\begin{equation}
\omega _1 =\frac{\sqrt{\omega _c ^2-2\omega _z ^2}}{2}.
\end{equation}
For more than one trapped ion besides the center-of-mass motion also other motional modes play a role. Two identical ions in a Penning trap are cooled with lasers until they form a crystalline structure. In our experiment, the orientation of the crystal along the magnetic field is energetically favorable. For this configuration, the mode frequencies in the radial plane are given by \cite{Guti2019b} 
\begin{equation}
\Omega _{c',m}^{\pm}=\frac{\omega _c}{2}\left [1\pm \sqrt{1-2\left (\frac{\Omega _z^{\pm}}{\omega _c}\right )^2}\,\,\right], \label{eq:omega_r}
\end{equation}
being
\begin{equation}
\Omega _z^-=\omega_z, \,\,\,\, \Omega _z^+=\sqrt{3}\omega_z \label{eq:omega_z}
\end{equation}
the eigenfrequencies in the axial direction.  One can define as well
\begin{equation}
\omega_1^{\pm}=\frac{\sqrt{\omega _c ^2-2(\Omega _z ^{\pm})^2}}{2}. \label{eq:w_1_crystal}
\end{equation}
The quantum Hamiltonian of the crystal, up to the ground state energy, can be written in terms of creation and anhilation operators \cite{Cerr2021}
\begin{align}
\nonumber H  &=\hbar\Omega_{z}^+a_{+}^{\dagger}a_{+} +\hbar\Omega_{z}^-a_{-}^{\dagger}a_{-}\\
\nonumber &+ \hbar\Omega_{c'}^+b_{c',+}^{\dagger}b_{c',+} + \hbar\Omega_{c'}^-b_{c',-}^{\dagger}b_{c',-}\\
& - \hbar\Omega_{m}^+b_{m,+}^{\dagger}b_{m,+} -\hbar\Omega_{m}^-b_{m,-}^{\dagger}b_{m,-},\label{eq:ham}
\end{align}
with the annihilation operator of the axial modes 
\begin{equation}
a_{\pm} =\frac{1}{\sqrt{2\hbar}}\left(\sqrt{m_{s}\Omega_z^{\pm}}z_{\pm}+i\sqrt{\frac{1}{m_{s}\Omega_z^{\pm}}}p_{z,\pm}\right),\label{eq:anhilation_axial}
\end{equation}and the annihilation operators of the radial modes: modified-cyclotron normal modes $b_{c',+}$, $b_{c',-}$ and magnetron normal modes $b_{m,+}$, $b_{m,-}$ given by
\begin{equation}
b_{k,\pm}=\frac{1}{\sqrt{2\hbar}}\left(\sqrt{\frac{m\omega_1^{\pm}}{2}}k_{\pm}+i\sqrt{\frac{2}{m\omega _1^{\pm}}}p_{k,\pm}\right). \label{eq:b}
\end{equation}

The minus sign in the last two terms in Eq.~(\ref{eq:ham}) establishes the instability of the magnetron modes which needs to be overcome in the cooling process.

\subsection{Cooling of externally produced ions}

\begin{figure}[t]
\centering\includegraphics[width=1.0\linewidth]{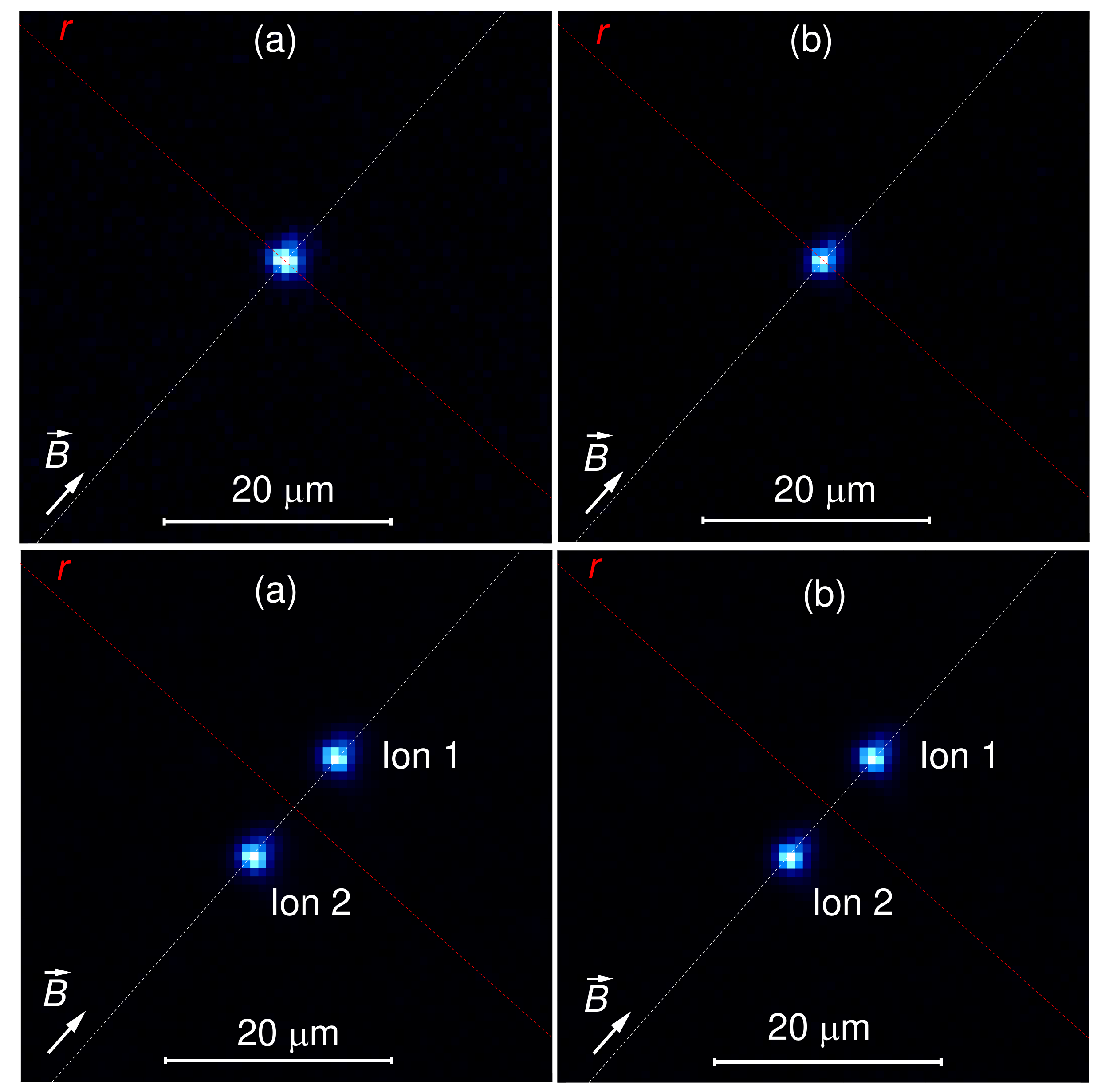}
\vspace{-0.5cm}
\caption{Single $^{40}$Ca$^+$ ion and balanced crystal in the open-ring Penning trap when the ions are created in the Paul trap, extracted, transferred, captured and laser-cooled in the Penning trap. 
Axialization is applied in (b). See text for further details.} \label{fig:crystal}
\end{figure}

External ion production has improved the vacuum conditions in the Penning trap. It is also mandatory in order to implement an unbalanced crystal \cite{Guti2019b} and to perform motional quantum metrology in a Penning trap \cite{Cerr2021} or other laser-based experiments (see e.g. \cite{Guti2019}). 
Figure~\ref{fig:crystal} shows single laser-cooled ions and balanced ion crystals when $V_{\scriptsize{\hbox{EC}}}=180$~V, $V_{\scriptsize{\hbox{CE}}}=168$~V and $V_{\scriptsize{\hbox{RE}}}=150$~V, corresponding to a measured value of $\omega _z = \Omega _z^- =2\pi \times 333$~kHz. The images are collected with an Electron-Multiplying Charge-Coupled-Device (EMCCD) camera and are the average of ten acquisitions each with a time window of one~second. The magnification of the optical system is around $18\times$ and is corrected for aberrations. The radial laser beams are directed perpendicular to the magnetic-field ($z$) axis in Fig.~\ref{fig:crystal}. Radial cooling is accomplished for both, the single ion and the two-ion crystal, displacing the radial beams with respect to the trap center along the $r$ line (Fig.~\ref{fig:crystal}a), and combining this with the application of a quadrupolar radiofrequency (RF) field at $\omega_c=2\pi\times 2.689370$~MHz (Fig.~\ref{fig:crystal}b) to exchange energy between the radial motions or modes, with amplitudes of 100~mV$_{\scriptsize{\hbox{pp}}}$ and 25~mV$_{\scriptsize{\hbox{pp}}}$ for the single ion and the two-ion crystal, respectively.

\begin{figure}[t]
\centering\includegraphics[width=0.9\linewidth]{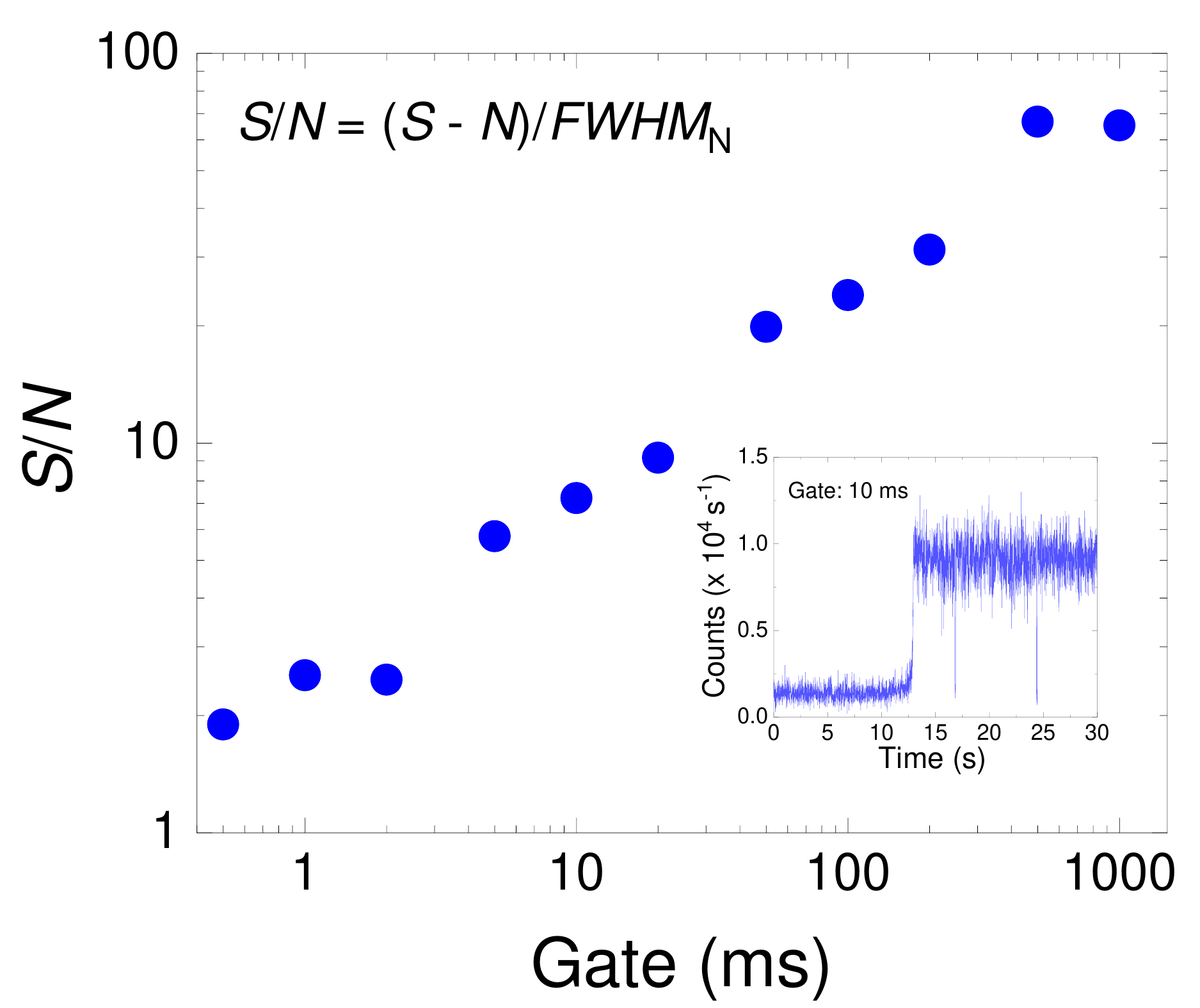}
\vspace{-0.4cm}
\caption{397-nm photons signal-to-noise ratio from a single laser-cooled $^{40}$Ca$^+$ ion in the open-ring Penning trap as a function of the acquisition gate in the photomultiplier tube. The inset shows the PMT signal for a time window of 30~s and a gate of 10~ms. The {\it FWHM}$_{\scriptsize{\hbox{N}}}$ is obtained from the standard deviation of the background signal.} \label{fig:PMT}
\end{figure}

The EMCCD image provides more sensitive information of the system after Doppler cooling, compared to measuring the number of photons with a photomultiplier tube (PMT). For example, the projections in the radial direction of the images in the upper panel of Fig.~\ref{fig:crystal} yield $\Delta r _{\scriptsize{\hbox{FWHM}}}=2.25(17)$~$\mu$m and $\Delta r_{\scriptsize{\hbox{FWHM}}}=1.87(9)$~$\mu$m for (a) and (b), respectively, while with the PMT one cannot observe differences in the number of photons. However, the time for cooling can be better monitored with the PMT. Figure~\ref{fig:PMT} shows the photons signal-to-noise ratio for a single laser-cooled $^{40}$Ca$^+$ ion recorded with a PMT simultaneously with the image in the EMCCD camera. Using the PMT, it is possible to utilize acquisition windows as short as 500~$\mu$s. 

We have investigated the time $t$ needed to cool a single $^{40}$Ca$^+$ ion and to form a $^{40}$Ca$^+$-$^{40}$Ca$^+$ crystal when these are injected externally. Under perfect vacuum, this time $t$ can be estimated using the model described in Ref.~\cite{Wese2007}. Defining $s=I/I_s$ as the saturation parameter and $E_0=\hbar \Gamma \sqrt{1+s}/2$ with $\Gamma =$$\Gamma$($^{40}$Ca$^+$)$=2\pi \times 21.6$~MHz (S$_{1/2}\rightarrow$P$_{1/2}$ transition), $t$ is related to the initial trapped-ion's energy $E$ by
\begin{equation}
t=\frac{4}{3}t_0\sqrt{r}\left (\frac{E}{E_0}\right )^{3/2} \label{eq:cooling_time}
\end{equation}
\noindent with $t_0=(1+s)/(s\Gamma /2)$, and $r=(\hbar K)^2/(2mE_0)$ with $K$ the photon's wave number. 
In a series of 17 measurements with a single ion, we obtain $t=165(64)$~seconds.   
We assume that the main energy is in the axial motion as inferred from the measurements in Sec.~\ref{sectwo}.

Taking $s=3.8$, it is possible to deduce a lower limit for $E$ of 4.2(2.3)~eV, which does not coincide with the value obtained in Sec.~\ref{sectwo}, for the main ion distribution. We can attribute the energy difference to the interaction between the ions and the residual-gas atoms. To prove that, we have blocked the cooling laser beams after injection of a single ion in the Penning trap for time intervals $t_{\scriptsize{\hbox{B}}}$ ranging from  $0$ to $200$~s in a total of 16~measurements. The time $t$ needed to cool one ion to the Doppler limit converges to $t_{B}$, allowing the determination of a decay-time constant of $225$~s, that only depends on collisions between the ions and the gas atoms, and therefore it is applicable to any ion species. The cooling time for an unbalanced crystal  will be thus shorter in the presence of residual gas atoms and appropriate for the experiments envisaged since the trapping time is above one hour.

\subsection{The two-ion crystal in a high magnetic field}\label{Ax}
\begin{figure}[t]
\centering\includegraphics[width=1.0\linewidth]{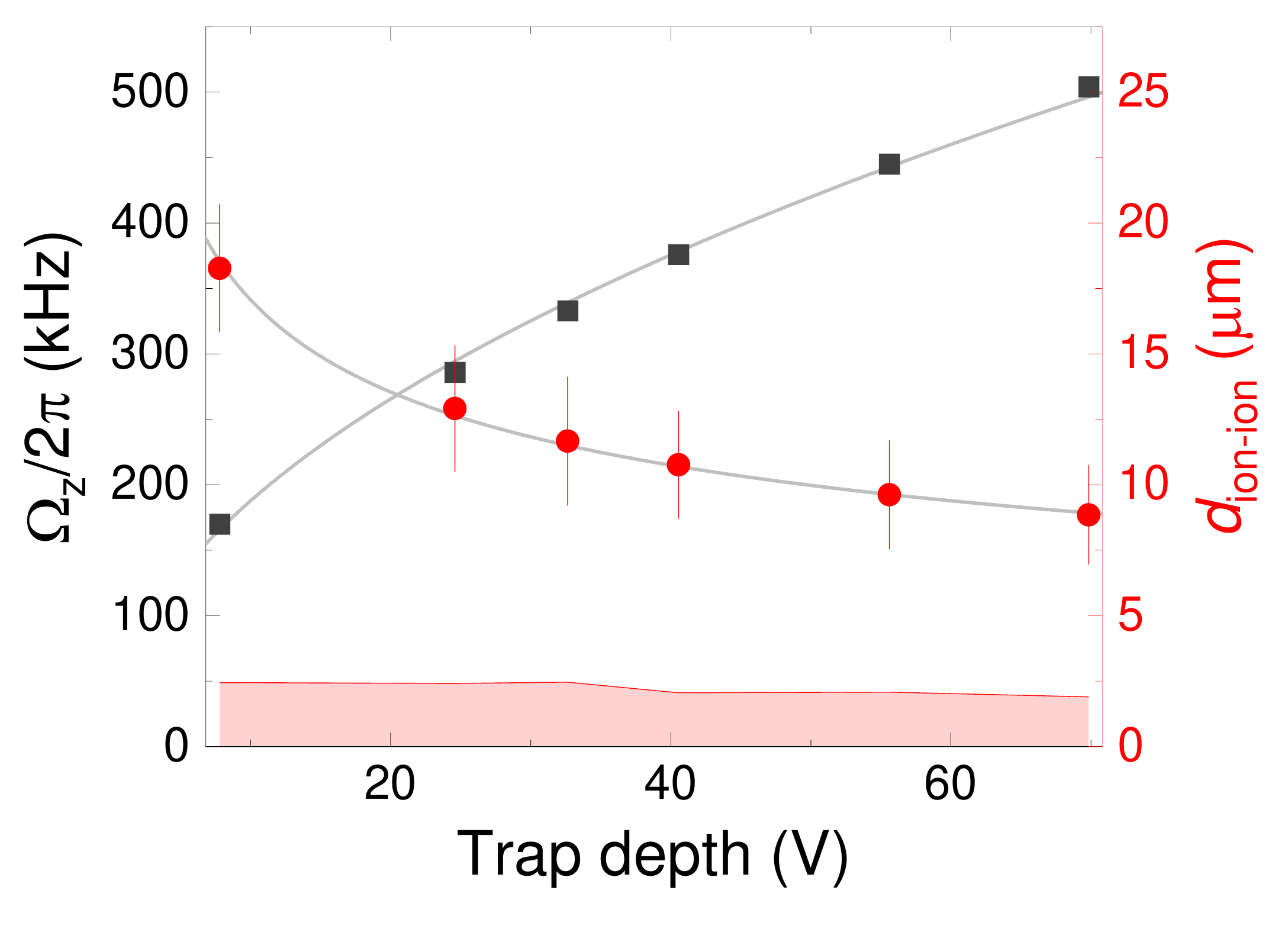}
\vspace{-0.8cm}
\caption{Evolution of the common axial frequency measured (solid squares) and optically-determined ion-ion distance (solid circles) versus trap depth. The solid lines are fits to the data points. The fitting function for the frequency data is proportional to the square root of the trap depth while that for $d_{\scriptsize{\hbox{ion-ion}}} $ is inversely proportional to the cube root. The values of the axial frequencies are listed in the third column in Tab.~(\ref{table_freq}). The uncertainties of the calculated ion-ion distance represent 2$\sigma$ of the projection from the photon distribution of a single ion. In this representation, one of the ions in the crystal is at zero and the red-shaded area displays the 2$\sigma$ range.}\label{fig:trap}
\end{figure}

By tuning the trap we have covered a frequency range from $\Omega _z^-/2\pi=170$-$504$~kHz. Figure~\ref{fig:trap} shows the axial-frequency values for different potentials applied to the trap electrodes. It also indicates the ion-ion distances from calculations including the uncertainties from the projection (Gaussian distributions) of the EMCCD image in the axial direction. Table~(\ref{table_freq}) shows the frequency values of the motional modes for the $^{40}$Ca$^+$-$^{40}$Ca$^+$ crystals built in this work, corresponding to the same data points appearing in Fig.~\ref{fig:trap}. The large strength of the magnetic field allows us to reach large values of $\Omega _z^{\pm}$, maintaining very large $\Omega _{c'}^{\pm}$. The main issue to be overcome after Doppler cooling is the magnetron motion with large  $\langle n_{m}\rangle$, where the letter $n$ represents the phonon number. Larger values of $\omega _{c'}$ and $\omega _{m}$ due to the high magnetic field imply lower $\langle n_{c'}\rangle $ and $\langle n_{m}\rangle$, further reducing $\langle n_{m}\rangle$ when applying axialization \cite{Hrmo2019}. 

The large magnetic field also improves the efficiency of axialization. The introduction of a quadrupolar RF driving field $A(t)$ in the radial plane has generally the Hamiltonian form 
\begin{equation}
 H_{\scriptsize{\hbox{ax}}}=A(t)\sum_{j=1,2}\left(x_j^2-y_j^2\right)\label{eq:Ax},
\end{equation}
where $A(t)$ is in units of energy per area, and the index $j$ refers to each of the ions. This term is invariant under the normal mode transformation of $u_{\pm}=(u_1\pm u_2)/\sqrt{2}$ with $u=x,y$ yielding
\begin{equation}
 H_{\scriptsize{\hbox{ax}}}=A(t)\sum_{j=\pm}\left(x_j^2-y_j^2\right).
\end{equation}
Further, after expressing this in terms of creation and annihilation operators and transforming into modified-cyclotron and magnetron modes, we obtain
\begin{align}
 H_{\scriptsize{\hbox{ax}}}=&A(t)\sum_{j=\pm}2 X^2_j\left(ib_{c',j}b_{m,j}^\dagger+H.c.\right)+\nonumber
 \\&A(t)\sum_{j=\pm}X^2_j\left(b_{c',j}^2+b_{c',j}^{\dagger 2}-b_{m,j}^2-b_{m,j}^{\dagger 2}\right),
\end{align}
where $X_j=\sqrt{2\hbar/(m\omega_1^j)}$, with $\omega_1^{\pm}$ defined in Eq.~(\ref{eq:w_1_crystal}) and  $b_{k,\pm}$ in Eq.~(\ref{eq:b}).
\begin{table}[t]
\caption{Motional modes frequencies of a $^{40}$Ca$^+$-$^{40}$Ca$^+$ crystal for different trap depths in a magnetic field of 7~tesla. The common axial frequencies have been probed using an external dipolar field in the axial direction. They are plotted for the different trap configurations in Fig.~\ref{fig:trap}. The radial frequencies are calculated using Eq.~(\ref{eq:omega_r}), and the stretch axial frequency using Eq.~(\ref{eq:omega_z}). $V_{\hbox{\scriptsize{TD}}}$ represents the trap depth in volts.}
\centering
\renewcommand{\arraystretch}{1.5}
\setlength{\tabcolsep}{1.5mm}

	\begin{tabular}{ccccccc}
			\hline\hline
$\Omega _{c'}^-/2\pi$ &  $\Omega _{c'}^+/2\pi$  & $\Omega _{z}^{-}/2\pi$  & $\Omega _{z}^+/2\pi$ & $\Omega _{m}^-/2\pi$&  $\Omega _{m}^+/2\pi$& $V_{\hbox{\scriptsize{TD}}}$\\
(MHz)&  (MHz)  & (kHz)  & (kHz)& (kHz)& (kHz) & (V)\\
\hline \hline
			2.684&  2.673&  170&  294&  5.4&  16& 7.8\\
                         2.674&  2.643&  286&  495&  15&  46& 24.6\\
                         2.669&  2.626&  333&  577&  21&  63& 32.6\\
                         2.663&  2.608&  376&  651&  27&  81& 40.6\\
                         2.652&  2.574&  445&  771&  37&  115& 55.6\\
                         2.641&  2.539&  504&  873&  48&  150& 69.8\\
                  	\hline \hline 
	\end{tabular} \label{table_freq}
\end{table}
A driving of the form $A(t)=A_{\scriptsize{\hbox{ax}}}\sin(\omega_c t)$ allows for the introduction of a Rotating Wave Approximation (RWA) in the interaction picture with respect to $H$, thus
\begin{equation}
 H'_{\scriptsize{\hbox{ax}}}=A_{\scriptsize{\hbox{ax}}}\sum_{j=\pm}X^2_j\left(ib_{c',j}b_{m,j}^\dagger+H.c.\right)
\end{equation}
which increases its validity with increasing magnetic field $B$. This term corresponds to a phonon exchange between magnetron and modified-cyclotron modes. As shown below, this is the reason why both modes end up sharing the same amount of phonons after axialization. Due to the smaller frequency, this means the magnetron motion will end up with a much smaller temperature than the modified-cyclotron motion. For axialization, we start considering the modified-cyclotron and magnetron motion of a single ion. The effect of cooling may be expressed in terms of the following master equation in the interaction picture with respect to Eq.~(\ref{eq:ham})
\begin{align}
 \dot\rho=&-iA_{\scriptsize{\hbox{ax}}} X^2\left[\left(ib_{c'}b_{m}^\dagger+H.c.\right),\rho\right]\nonumber\\
+ &\sum_{j=c',m}R^h_j\left(b_{j}^\dagger\rho b_{j}-\frac{1}{2}b_{j}b_{j}^{\dagger}\rho-\frac{1}{2}\rho b_{j}b_{j}^{\dagger}\right)\nonumber\\
+ &\sum_{j=c',m}R^c_j\left(b_{j}\rho b_{j}^\dagger-\frac{1}{2}b_{j}^\dagger b_j\rho-\frac{1}{2}\rho b_{j}^\dagger b_{j}\right).
\end{align}
where $R_j^c$ and $R_j^h$ represent cooling and heating rate of the corresponding mode, respectively. These rates correspond respectively to the absorption and emission strengths of the optical transition at the detuning of the mode's frequency \cite{Cira1992}. Without axialization ($A_{\scriptsize{\hbox{ax}}}=0$), each motion cools at a constant rate $R_j=R^c_j-R^h_j$ and reaches a final phonon number $\left\langle n_j\right\rangle_f=R^h_j/R_j$. With axialization, assuming the cooling rates are still constant, the dynamics are described by the equations
\begin{align}
 &\frac{d}{dt}\left\langle n_{c'}\right\rangle=-A_{\scriptsize{\hbox{ax}}}  X^2\left\langle c \right\rangle - R_{c'}\left\langle n_{c'}\right\rangle +R^h_{c'}, \nonumber\\
 &\frac{d}{dt}\left\langle n_{m}\right\rangle=A_{\scriptsize{\hbox{ax}}}  X^2\left\langle c \right\rangle - R_{m}\left\langle n_{m}\right\rangle +R^h_{m}, \nonumber\\
 &\frac{d}{dt}\left\langle c \right\rangle=-\frac{R_{c'}+R_m}{2}\left\langle c \right\rangle + 2A_{\scriptsize{\hbox{ax}}} X^2 \left(\left\langle n_{c'}\right\rangle-\left\langle n_m\right\rangle\right), \label{eq:ax_single}
\end{align}
where  $c=b_{c'}b_m^\dagger+b_{c'}^\dagger b_m$. This term works towards reducing the phonon difference between both motions. The steady state occupation under axialization then becomes
\begin{align}
 &\left\langle n_{c'}\right\rangle_f^{ax}=\dfrac{4A_{\scriptsize{\hbox{ax}}} ^2X^4\left\langle n_{ax}\right\rangle+R_{c'}R_m\left\langle n_{c'}\right\rangle_f}{4A_{\scriptsize{\hbox{ax}}} ^2X^4+R_{c'}R_m},\\
 &\left\langle n_{m}\right\rangle_f^{ax}=\dfrac{4A_{\scriptsize{\hbox{ax}}} ^2X^4\left\langle n_{ax}\right\rangle+R_{c'}R_m\left\langle n_{m}\right\rangle_f}{4A_{\scriptsize{\hbox{ax}}} ^2X^4+R_{c'}R_m},
\end{align}
where we have defined
\begin{equation}
\left\langle n_{ax}\right\rangle =\frac{R_{c'}\left\langle n_{c'}\right\rangle_f+R_m\left\langle n_{m}\right\rangle_f}{R_{c'}+R_m},
\end{equation}
which is the phonon number both motions end up sharing under strong axialization $A_{\scriptsize{\hbox{ax}}} X^2\gg \sqrt{R_{c'}R_m}/2$:
\begin{equation}
\left\langle n_{c'}\right\rangle_f^{ax}=\left\langle n_m\right\rangle_f^{ax}=\left\langle n_{ax}\right\rangle. \label{eq:ax_final}
\end{equation}
This holds true provided $R_j^c$ and $R_j^h$ are constant \cite{Cira1992}. This might not be the case when $A_{\scriptsize{\hbox{ax}}}$ is too large \cite{Scha2018}.

\begin{figure}[t]
\centering\includegraphics[width=1.05\linewidth]{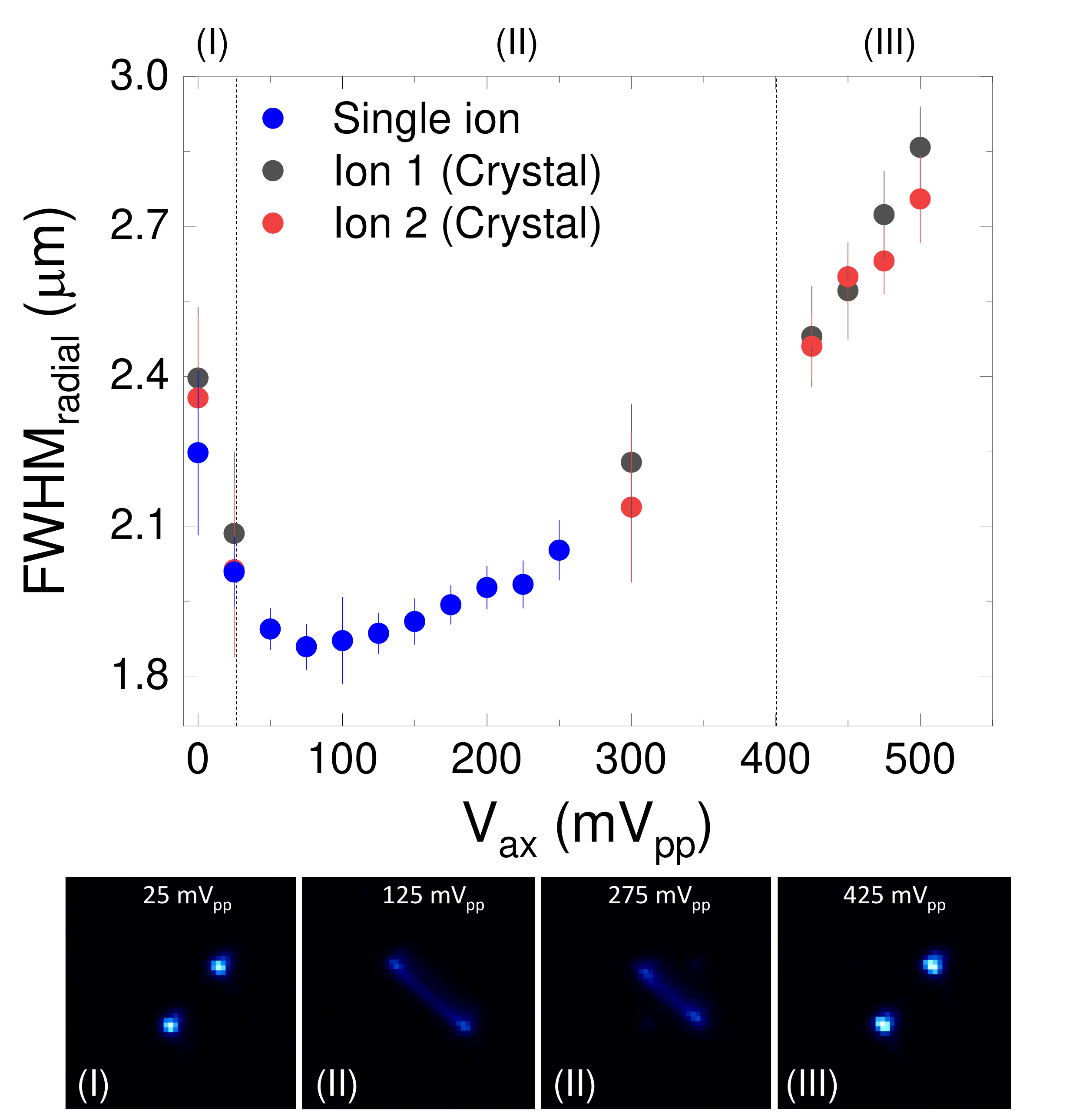}
\vspace{-0.5cm}
\caption{Evolution of the $\langle ${\it FWHM}$\rangle_{\hbox{\scriptsize{radial}}}$ as a function of the axialization amplitude for one ion and for the two ions forming a balanced crystal. The crystal is stable along the axial direction in regions (I) and (III). Each data point is the average of ten consecutive measurements. In region (II) the crystal is oriented most of the time in the radial plane. In this particular series of measurements, for $V_{\scriptsize{\hbox{ax}}}=300$~mV$_{\scriptsize{\hbox{pp}}}$, the crystal is oriented along the axial direction most of the time. Some characteristic images from the different regions are shown.}
\label{evolution}
\end{figure}
Under the best conditions observed while tuning the radial laser beams, the improvement introduced by applying axialization has been clearly observed from the projections of images as those shown in Fig.~\ref{fig:crystal}, along the radial direction. We have analyzed a set of subsequent measurements and the mean values of the {\it FWHM} from the single-ion projections in the radial direction are $\langle ${\it FWHM}$\rangle=2.25(17)$~$\mu$m when axialization is not applied, and $\langle ${\it FWHM}$^{\scriptsize{\hbox{ ax}}}\rangle=1.87(9)$~$\mu$m when $V_{\scriptsize{\hbox{ax}}}=100$~mV$_{\scriptsize{\hbox{pp}}}$. Note that $A_{\scriptsize{\hbox{ax}}} \propto V_{\scriptsize{\hbox{ax}}}$ and $V_{\scriptsize{\hbox{ax}}}$ is the amplitude of the quadrupolar field applied (V$_{\scriptsize{\hbox{pp}}}$). The uncertainties correspond to one standard deviation. The position of the ion along the radial direction does not change ($\Delta r =10(80)$~nm). This shows a significant improvement when applying axialization. The values of $V_{\scriptsize{\hbox{ax}}}$ optimal for the single-ion case are in the range 50-150~mV$_{\scriptsize{\hbox{pp}}}$, observing for $V_{\scriptsize{\hbox{ax}}}=150$-$250$~mV$_{\scriptsize{\hbox{pp}}}$ slightly worse performance increasing with $V_{\scriptsize{\hbox{ax}}}$, although still better compared to the case where $V_{\scriptsize{\hbox{ax}}}=0$~mV$_{\scriptsize{\hbox{pp}}}$, as shown in Fig.~\ref{evolution}. This behaviour does not follow the ideal case of  Eq.~(\ref{eq:ax_final}), where $R_j^c$ and $R_j^h$ were kept constant and unaffected by the strength of $A_{\scriptsize{\hbox{ax}}}$. The {\it FWHM} in the axial direction remains the same from $V_{\scriptsize{\hbox{ax}}}=0$ to $V_{\scriptsize{\hbox{ax}}}=250$~mV$_{\scriptsize{\hbox{pp}}}$.

Axialization of the common and breathing mode of the crystal in the radial plane should be accomplished with the same RF of the field since $\Omega _{c'}^{\pm}+\Omega _{m}^{\pm} \approx \omega _c$ (Tab.~(\ref{table_freq})). In this respect, the common radial modes can be treated similarly to the radial motions of a single-ion (Eq.~(\ref{eq:ax_single})). However, this might not be the case for the breathing modes, which although following the same equation, are expected to differ in the cooling rates.

The values of the {\it FWHM} from the projections in the radial directions of the images of the two ions in Fig.~\ref{fig:crystal} are $\langle ${\it FWHM}$\rangle=2.40(14)$~$\mu$m and $\langle ${\it FWHM}$^{\scriptsize{\hbox{ ax}}}\rangle=2.09(16)$~$\mu$m for ion 1, and $\langle${\it FWHM}$\rangle=2.36(16)$~$\mu$m and $\langle${\it FWHM}$^{\scriptsize{\hbox{ ax}}}\rangle=2.01(17)$~$\mu$m for ion 2. For this measurement $V_{\scriptsize{\hbox{ax}}}=25$~mV$_{\scriptsize{\hbox{pp}}}$. For larger values of $V_{\scriptsize{\hbox{ax}}}$ the two ions align on the radial plane as shown for two cases in Fig.~\ref{evolution}. The crystal becomes stable again for $V_{\scriptsize{\hbox{ax}}}\geq 425$~mV$_{\scriptsize{\hbox{pp}}}$. In this region (III) in Fig.~\ref{evolution}, the position of the ion crystal moves by about 1~nm/mV$_{\scriptsize{\hbox{pp}}}$. This might be due to deviations from the ideal quadrupolar field arising from mechanical imperfections or small misalignments between electrodes. This hypothesis is reinforced by observing the increasing trend in  $\langle ${\it FWHM}$\rangle_{\hbox{\scriptsize{radial}}}$ in Fig.~\ref{evolution}.

While for a single ion these possible effects do not counteract the improvement due to axialization when $V_{\scriptsize{\hbox{ax}}}>25$~mV$_{\scriptsize{\hbox{pp}}}$, this is not the case for two ions (region (II) in Fig.~\ref{evolution}). Although $\Omega _{c'}^{\pm}+\Omega _{m}^{\pm} \approx \omega _c$, the two-ion crystal along the magnetic field axis can be perturbed more easily as observed in Fig.~\ref{evolution}  and thus, the crystalline structure will be more sensitive to possible deviations from the ideal quadrupolar field. Our theoretical approach is based on a harmonic approximation around the axial potential minima of each ion. This approximation is valid as long as the axial angular momentum $L_z$ of the crystal remains negligible. Beyond a certain threshold of $L_z$, the potential minima align on the radial plane instead, rendering the axial configuration unstable and the radial one stable. Due to their axial symmetry, the bare Penning trap potentials conserve $L_z$ and cooling contributes to its reduction. Therefore, the axial configuration is stable for most cooling scenarios. Nevertheless, this is not necessarily the case in the presence of axialization. Axialization breaks the axial symmetry and therefore has the potential to increase $L_z$. The breathing mode, being the only one affected by the non-linear Coulomb repulsion, is most susceptible to this effect. Together with the possibility of small misalignments mentioned above, this intriguing behaviour is the result of non-linear effects that establish the ground for further investigation.

\section{Conclusions and Outlook}
In this publication we have shown a single laser-cooled ion and a two-ion crystal in a 7-tesla Penning trap when the ion species ($^{40}$Ca$^+$) is produced by photoionization in a Paul trap in the absence of a magnetic field, extracted, transported and captured in-flight in the open-ring Penning trap. Balanced crystals have been formed for several trap configurations allowing a range of common axial frequencies of a few hundred kHz. The energy of the ions with respect to the minimum of the trap potential is a few electronvolts. The effect of having a high magnetic-field strength has been also analyzed and we have found two advantages. First, the high-magnetic field reduces the contribution of residual terms counteracting the cooling process of the radial motion under axialization. Second, the same field will yield a smaller mean phonon number  in the magnetron motion in the axialization process if the cooling rate of the modified-cyclotron motion (single ion) or modes (two-ion crystal) are much larger than the magnetron ones. Although the advantage of axialization has been demonstrated for a single laser-cooled ion, the results are not optimal for the balanced crystal, which is more sensitive to perturbations, for example arising from misalignements of the electrodes  and with cooling rates for the common (like the single-ion case) and stretch modes that can be different. In order to achieve similar cooling results for both modes, different optimal amplitudes ($V_{\scriptsize{\hbox{ax}}}$) should be applied for each. This is not possible, since both modes require the same axialization frequency $\omega _c$. This problem will be lifted in the case of unbalanced crystals, since the relationship $\Omega _{c'}^{\pm}+\Omega _{m}^{\pm} \approx \omega _c$ does not hold.

The performance of the system, obtained from the fluorescence images, suggests that Doppler cooling of an unbalanced crystal made of a $^{40}$Ca$^+$ ion and any other charged particle is feasible overcoming the instability of the magnetron motion and allowing to use the platform for motional metrology or other laser-based or laser-assisted experiments. Motional metrology has been presented proposing $^{232}$Th$^+$ as the target ion \cite{Cerr2021}. The production and injection of this ion species is currently under investigation. When $\omega _z/2\pi=333$~kHz, the frequencies of the motional modes of the crystal $^{232}$Th$^+$-$^{40}$Ca$^+$ using the linear approximation \cite{Guti2019b}, that is valid in the quantum regime, are $\Omega _{c'}^-/2\pi=2.647$~MHz, $\Omega _{c'}^+/2\pi=416$~kHz, $\Omega _{m}^-/2\pi=68$~kHz, $\Omega _{m}^+/2\pi=21$~kHz, $\Omega _{z}^-/2\pi=165$~kHz and $\Omega _{z}^+/2\pi=482$~kHz. This requires two quadrupolar fields for axialization, one at a frequency $\Omega _{c'}^{-}+\Omega _{m}^{-} \approx 2\pi \times 2.715$~MHz and another with $\Omega _{c'}^{+}+\Omega _{m}^{+} \approx 2\pi \times 437$~kHz. This will allow applying two quadrupolar fields simultaneously and observing the cooling of the two modes. In any case, in order to measure the final temperature and to carry out experiments of quantum metrology with the unbalanced crystal, the S$_{1/2}\rightarrow $D$_{5/2}$ (clock) transition in $^{40}$Ca$^+$ \cite{Chwa2009} needs to be addressed to bring the crystal into the ground state and to measure accurately the motional frequencies of its common and breathing modes.

\section*{Acknowledgement}
We acknowledge support from the Spanish MICINN through the project PID2019-104093GB-I00/AEI/10.01339/501100011033 and contract PTA2018-016573-I, and from the Andalusian Government through the project P18-FR-3432 and Fondo Operativo FEDER A-FQM-425-UGR18, from the Spanish Ministry of Education through PhD fellowship FPU17/02596, and from the University of Granada "Plan propio - Programa de Intensificaci\'on de la Investigaci\'on", project PP2017-PRI.I-04 and "Laboratorios Singulares 2020". The construction of the facility was supported by the European Research Council (contract number 278648-TRAPSENSOR), projects FPA2015-67694-P and FPA2012-32076, infrastructure projects UNGR10-1E-501, UNGR13-1E-1830 and EQC2018-005130-P (MICINN/FEDER/UGR), and IE-5713 and IE2017-5513 (Junta de Andaluc\'ia-FEDER). JC acknowledges support from the Spanish MICINN ("Beatriz Galindo" Fellowship BEAGAL18/00081).

\appendix

\section{Ions' temperature in the Coulomb crystal}\label{coulomb}
\begin{figure}[t]
\centering\includegraphics[width=1.0\linewidth]{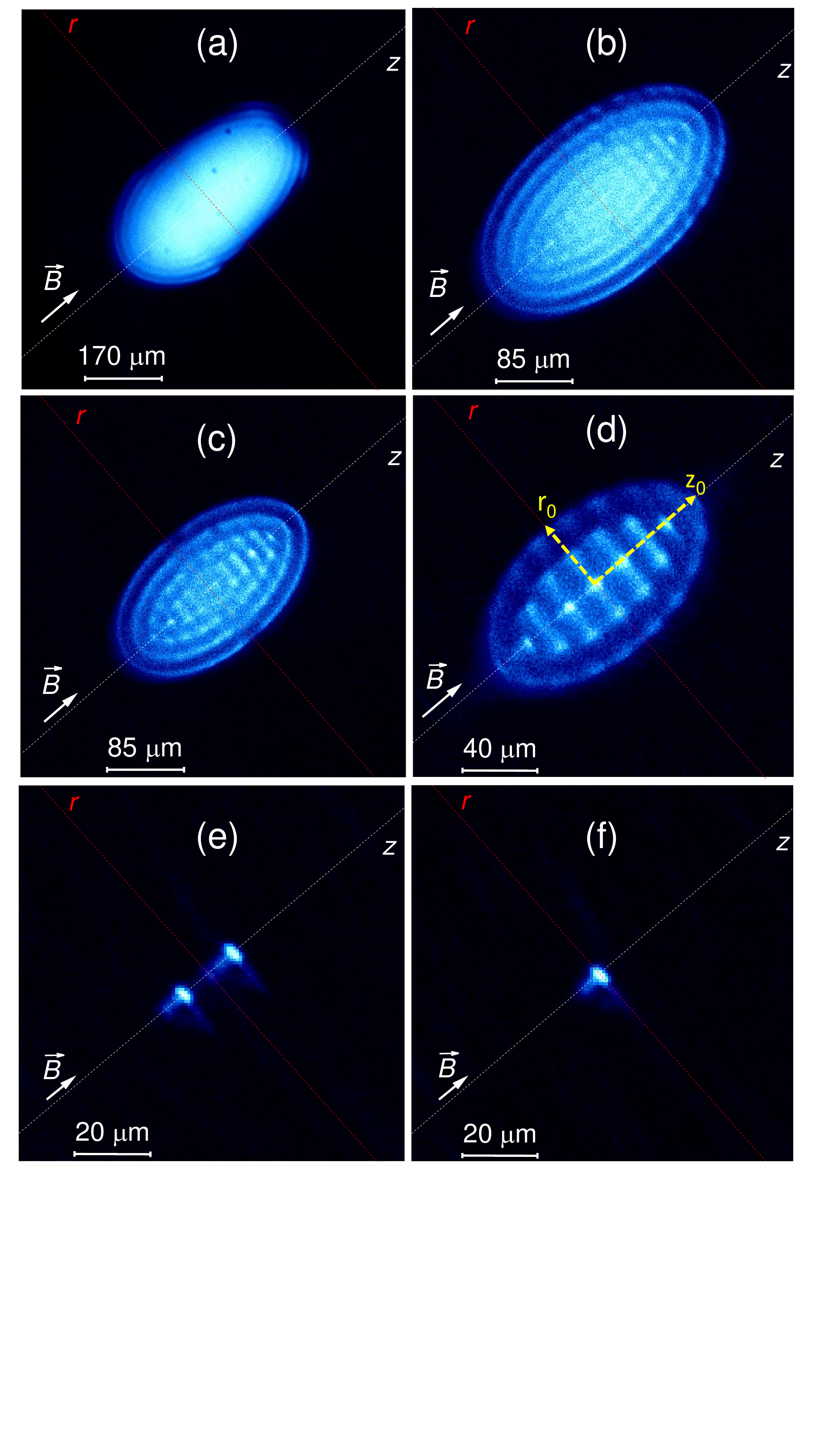}
\vspace{-3.6cm}
\caption{Images of ICCs in the 7-tesla Penning trap formed by different amount of ions. The crystal structure becomes clearly visible from (b). In all these cases the ions were created inside the Penning trap. The open-ring Penning trap was operated with an axial oscillation frequency for a single $^{40}$Ca$^+$ ion in the center of the trap $\omega_z=2\pi \times 170$~kHz. } \label{figure_crystals}
\end{figure}
\begin{figure}[t]
\centering\includegraphics[width=1.0\linewidth]{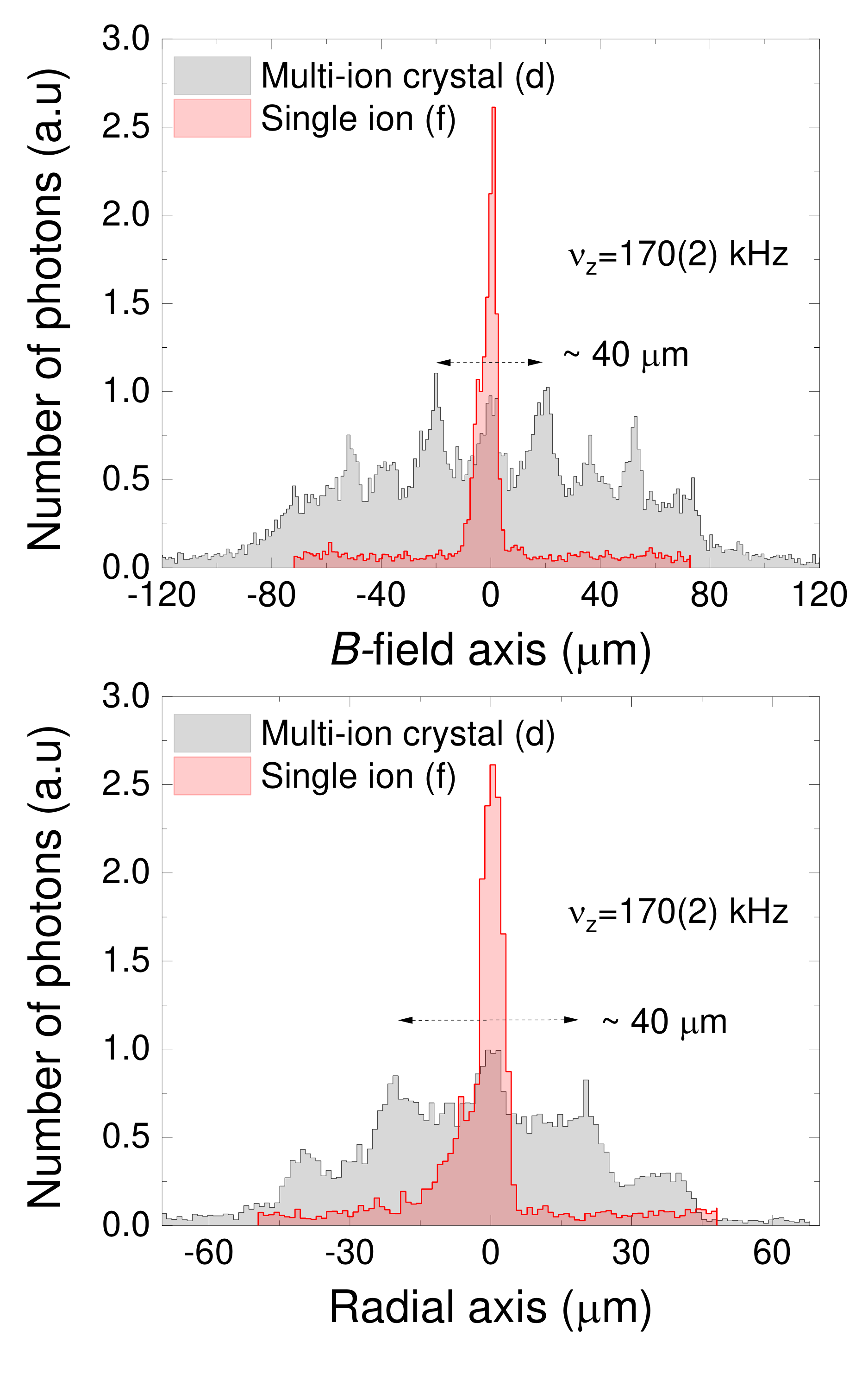}
\vspace{-1cm}
\caption{Top: Projections along the magnetic field axis of the crystal shown in Fig.~\ref{figure_crystals}(d) and that of the single ion in (f). Bottom: Projections along the radial direction of the crystal shown in Fig.~\ref{figure_crystals}(d) and that of the single ion in (f). The peaks are not fully resolved in this direction due to the rotation of the crystal forming an ellipsoid around the magnetic field axis.}\label{projection}
\end{figure}

Here we show Ion Coulomb Crystals (ICCs) and we use one of them to extract the ions' temperature. ICCs in Penning traps involving more ions have been observed and extensively studied at NIST-Boulder \cite{Boll1993,Brit2012} ($^9$Be$^+$ ions in 4.46~T) and Imperial College \cite{Mava2013} ($^{40}$Ca$^+$ ions in 1.86~T). The temperature of large \cite{Sawy2012} and small ICCs \cite{Stut2017} have been measured with these setups using techniques beyond Doppler cooling. In the experiments reported here, ICCs have been formed as shown in Fig.~\ref{figure_crystals} for ions created inside the 7-T Penning trap. The images are collected with the EMCCD camera and an acquisition time window of one~second. The magnification of the optical system here is around $\times15$. The crystal structure is clearly visible in Fig.~\ref{figure_crystals} from (b) down to a single ion in (f). The optical system was not corrected for aberrations. Figure~\ref{projection} shows the projections along the magnetic field axis and perpendicularly to it in the plane of the figure for the image in Fig.~\ref{figure_crystals}(d,f). 

Defining the ion density $n_0$, the ratio between $\omega _z$ and the plasma frequency \cite{Boll1993}
\begin{equation}
\omega _p^2=\frac{q^2n_0}{\epsilon_0 m}=2\omega_r(\omega_c-\omega_r)
\end{equation}
with $\epsilon _0$ being the vacuum permittivity and $\omega _r $ the rotation frequency of the crystal, is given as a function of the ratio $\alpha = z_0/r_0$ (Fig.~\ref{figure_crystals}(d)), by 
\begin{equation}
\frac{\omega_z^2}{\omega _p^2}=\frac{1}{\alpha^2-1}\left [\frac{u_p}{2}\hbox{ln}\left (\frac{u_p+1}{u_p-1}\right )-1\right ],
\end{equation}
with 
\begin{equation}
u_p=\frac{\alpha}{\sqrt{\alpha ^2-1}}.
\end{equation}
For the case shown in Fig.~\ref{figure_crystals}(d), $\omega _p=2\pi \times 386$~kHz, $\omega _r=2\pi \times 27.8$~kHz, and $n_0=1.3\times 10^8$~cm$^3$. The ion number is $\approx70$. From the plasma we can estimate the temperature $T$ observing e.g. Fig.~\ref{projection}, where it is possible to extract an average ion-ion distance of $\sim 20$~$\mu$m. This average distance is defined as Wigner-Seitz radius ($a_0$) in the Coulomb coupling parameter \cite{Thom2014}
\begin{equation}
\Gamma=\frac{e^2}{4\pi \epsilon_0a_0k_BT},
\end{equation}
where $k_B$ is the Boltzmann's constant. Taking $\Gamma =178$ (considering an infinite plasma) for a phase transition to occur \cite{Jone1996}, an upper limit for the temperature of $T\leq 5$~mK can be found.

\providecommand{\noopsort}[1]{}\providecommand{\singleletter}[1]{#1}%

\end{document}